\documentstyle[aps,twocolumn,floats]{revtex}
\begin{document}
\newcommand\1{$\spadesuit$}
\newcommand\2{$\clubsuit$}
\def\st{\scriptstyle}
\def\sst{\scriptscriptstyle}
\def\mco{\multicolumn}
\def\epp{\epsilon^{\prime}}
\def\vep{\varepsilon}
\def\ra{\rightarrow}
\def\ppg{\pi^+\pi^-\gamma}
\def\vp{{\bf p}}
\def\ko{K^0}
\def\kb{\bar{K^0}}
\def\al{\alpha}
\def\ab{\bar{\alpha}}
\def\be{\begin{equation}}
\def\ee{\end{equation}}
\def\bea{\begin{eqnarray}}
\def\eea{\end{eqnarray}}
\tighten
\draft
\twocolumn[\hsize\textwidth\columnwidth\hsize\csname
@twocolumnfalse\endcsname
\title{Termination of the Phase of Quintessence by Gravitational
Back-Reaction}

\author{Mingzhe Li, Wenbin Lin, Xinmin Zhang\\
Institute of High Energy Physics, Chinese Academy of Sciences, \\
P.O. Box 918-4, Beijing 100039, P.R. China,\\
e-mail: limz@hptc5.ihep.ac.cn, wblin@hptc5.ihep.ac.cn,
xmzhang@hptc5.ihep.ac.cn  \\
{\rm and}\\
Robert Brandenberger\\
Department of Physics, Brown University, Providence, RI 02912, USA.\\
e-mail: rhb@het.brown.edu.
}

\date{July 10, 2001}
\maketitle

\begin{abstract}

We study the effects of gravitational back-reaction in models
of Quintessence. The effective energy-momentum tensor with
which cosmological fluctuations back-react on the background
metric will in some cases lead to a termination of the phase
of acceleration. The fluctuations we make use of are
the perturbations in our present Universe. Their amplitude
is normalized by recent measurements of anisotropies
in the cosmic microwave background, their slope is taken
to be either scale-invariant, or characterized by a slightly
blue tilt. In the latter case, we find that the back-reaction
effect of fluctuations whose present wavelength is smaller
than the Hubble radius but which are stretched beyond the
Hubble radius by the accelerated expansion during the era
of Quintessence domination can become large. Since the
back-reaction effects of these modes oppose the acceleration,
back-reaction will lead to a truncation of the period of
Quintessence domination. This result impacts on the recent
discussions of the potential incompatibility between
string theory and Quintessence.

\end{abstract}

\vspace*{1cm}
]

\section{Introduction}

Evidence is increasing that the Universe is spatially flat and, at
the present time, accelerating. The best evidence for a
spatially flat Universe comes from the location of the
acoustic peaks of the spectrum of cosmic microwave anisotropies
(see e.g. \cite{Pierpaoli:2000xa} for a recent discussion).
Since dynamical mass determinations
from observations of
large-scale structure are converging \cite{LSS} to a mass density
far short of the critical density, the density required for
a spatially flat Universe, the difference must be due to
either a remnant small cosmological constant, or a new form
of matter which is not clustered gravitationally on the
scale of galaxy clusters, and has been
given various names, including {\it dark energy} and {\it quintessence},
the name we will adopt (see e.g. 
\cite{Wetterich:1988fm,Ratra:1988rm} for original
papers and \cite{Caldwell:2000wt} for a recent
review and detailed references). This conclusion is supported by the
recent supernova observations which yield Hubble diagrams which
directly support the evidence that the Universe is, at the
present time, accelerating \cite{Goobar:2000qy,Garnavich:1998th}.
Hence, Quintessence must
have an equation of state $p < - (1/3)\rho$, $p$ and $\rho$ denoting
the pressure and the energy density, respectively.

Most models of Quintessence proposed involve a new scalar field $Q$
which is taken to be homogeneous in space, and via its kinetic
and potential energy contributions to the energy-momentum tensor
$T_{\mu \nu}$ tuned to provide an equation of state
leading to accelerated expansion, beginning to dominate the matter
content of the Universe today. In many such models, acceleration will
continue forever.

It can easily be seen that in such models
there is an event horizon for every observer.  As has
recently been pointed out \cite{Fischler:2001yj,Hellerman:2001yi},
this leads to
a potential incompatibility between models of Quintessence
and string theory. It has been argued (see e.g. \cite{Witten:2001kn}
and references therein) that
in the presence of de Sitter horizons, string theory cannot
be defined since physical observables cannot be expressed
in terms of the S-matrix of string theory. As is obvious
\cite{Fischler:2001yj,Hellerman:2001yi},
the problem immediately carries over to models of Quintessence
postulated which today and, in many models, in all future, dominates
the matter content of the Universe (this conclusion can be avoided
in certain models with an exponential potential along ``tracker''
solutions \cite{Cline:2001nq,Kolda:2001ex}).

By adding more new fields, it is possible to solve (see e.g.
\cite{Halyo:2001fb,Gu:2001wp,Bento:2001yv})
this problem by constructing models in which the period
of Quintessence is terminated in a way which is analogous
to the termination of the period of inflation in hybrid inflation
\cite{Linde:1994cn} models. In this letter, we point out that
gravitational back-reaction of cosmological fluctuations provides a
mechanism which, in some models,
will lead to a termination of the phase of acceleration,
without the need to add new physics.

\section{Method and Qualitative Considerations}

The idea of gravitational back-reaction of cosmological fluctuations is
simple. In the presence of the fluctuations of the space-time metric
and of matter, the cosmological background evolves differently compared
to its evolution in the absence of perturbations
\cite{Mukhanov:1997ak,Abramo:1997hu}. This
effect is due to the nonlinearities of the Einstein field equations.
The effect can be characterized in terms of an effective energy-momentum
tensor $\tau_{\mu \nu}$ with which the fluctuations back-react on the
background metric.

In this paper, we will focus on the back-reaction effect of infrared
modes (modes with wavelength larger than the Hubble radius $H^{-1}(t)$,
where $H(t)$ is the Hubble expansion rate) on the evolution of a
Universe dominated by a Quintessence field. In a separate paper, we
will analyze the back-reaction effect by ultraviolet modes \cite{LLZ2}.

In models with accelerated expansion, the phase space of infrared modes
grows since the Hubble radius is decreasing in comoving coordinates, as
is the case in inflationary cosmology. Hence, we expect the back-reaction
effect of infrared modes to grow. In the following we will show that
in models with a sufficiently blue spectrum of primordial fluctuations,
the back-reaction terms will eventually dominate the field equations for
the effective background. Since the equation of state corresponding to
$\tau_{ij}$ opposes acceleration, back-reaction effects cannot be neglected
and are expected to terminate the accelerated expansion, or, in other
words, to screen the effects of the background Quintessence field.

We first give a brief review of gravitational
backreaction of a general scalar matter field $\varphi$
and then apply it to the period of
Quintessence domination.

The basic idea of gravitational backreaction (see e.g.
\cite{Brandenberger:1999su} for a review) is to expand the
Einstein equations to second order
in the perturbations, to assume that the first order terms satisfy
the equations
of motion for linearized cosmological perturbations (see e.g.
\cite{Mukhanov:1992me}
for a comprehensive review),
to take the spatial average of the remaining terms,
and to regard the resulting equations as equations for a new homogeneous
metric
$g_{\mu \nu}^{(0, br)}$ which includes the effect of the perturbations to
quadratic order:
\begin{eqnarray}\label{breq}
G_{\mu \nu}(g_{\alpha \beta}^{(0, br)}) \, =
\, 8 \pi G \left[ T_{\mu \nu}^{(0)} + \tau_{\mu \nu} \right]\, ,
\end{eqnarray}
where the effective energy-momentum tensor $\tau_{\mu \nu}$ of
gravitational back-reaction contains the terms resulting from
spatial averaging of the second order metric and matter perturbations:
\be \label{efftmunu}
\tau_{\mu \nu} \, = \, < T_{\mu \nu}^{(2)} - {1 \over {8 \pi G}}
G_{\mu \nu}^{(2)} > \, ,
\ee
where pointed brackets stand for spatial averaging, and the superscripts
indicate the order in perturbations.

In longitudinal gauge the perturbed metric can be written in the form
\be \label{metric}
ds^2 =  (1+ 2 \Phi) dt^2 - a(t)^2(1 - 2\Phi) \delta_{i j} dx^i dx^j  \, ,
\ee
where $a(t)$ is the cosmological scale factor. Provided there are
no linear contributions to the spatial off-diagonal terms in the
matter energy-momentum tensor, the above metric contains the full
information about scalar metric fluctuations.
The energy-momentum tensor for a scalar field is
\be
T_{\mu \nu }=\varphi _{,\mu }\varphi _{,\nu }-g_{\mu \nu }\left[ {\frac 12}
\varphi ^{,\alpha }\varphi _{,\alpha }-V(\varphi )\right] \,.
\ee
In this case, the spatial off-diagonal terms in $T_{\mu \nu}$ vanish.

By expanding the Einstein tensor and the above energy-momentum
tensor to second order in the metric and matter fluctuations
$\Phi$ and $\delta \varphi$, respectively, it can be shown that
the non-vanishing components of the effective back-reaction
energy-momentum tensor $\tau_{\mu \nu}$ become
\bea  \label{tzero}
\tau_{0 0} &=& \frac{m_{pl}^2}{8 \pi } \left[ + 12 H \langle \Phi \dot{\Phi} \rangle
- 3 \langle (\dot{\Phi})^2 \rangle + 9 a^{-2} \langle (\nabla \Phi)^2
\rangle \right]  \nonumber \\
&+& {1 \over 2} \langle ({\delta\dot{\varphi}})^2 \rangle + {1 \over 2}
a^{-2} \langle
(\nabla\delta\varphi)^2 \rangle  \nonumber \\
&+& {1 \over 2} V''(\varphi_0) \langle \delta\varphi^2 \rangle + 2
V'(\varphi_0) \langle \Phi \delta\varphi \rangle \quad ,
\eea
and
\bea  \label{tij}
\tau_{i j} &=& a^2 \delta_{ij} \left\{ \frac{m_{pl}^2}{8 \pi}
\left[ (24 H^2 + 16
\dot{H}) \langle \Phi^2 \rangle + 24 H \langle \dot{\Phi}\Phi \rangle
\right. \right.  \nonumber \\
&+& \left. \langle (\dot{\Phi})^2 \rangle + 4 \langle \Phi\ddot{\Phi}
\rangle - \frac{4}{3} a^{-2}\langle (\nabla\Phi)^2 \rangle \right]
+ 4 \dot{{
\varphi_0}}^2 \langle \Phi^2 \rangle  \nonumber \\
&+& {1 \over 2} \langle ({\delta\dot{\varphi}})^2 \rangle -
{1 \over 6} a^{-2} \langle
(\nabla\delta\varphi)^2 \rangle -
4 \dot{\varphi_0} \langle \delta \dot{\varphi}
\Phi \rangle  \nonumber \\
&-& \left. {1 \over 2} \, V''(\varphi_0) \langle \delta\varphi^2
\rangle + 2 V'( \varphi_0 ) \langle \Phi \delta\varphi \rangle
\right\} \quad .
\eea
All terms are quadratic in the fluctuations. The two-point functions
in the pointed brackets can be viewed classically as spatial averages
or quantum mechanically as equal-time two-point functions.

We will now apply these equations to Quintessence models and thus
replace $\varphi$ by $Q$.
Since we are interested in the back-reaction effect of infrared modes,
we can drop all terms involving spatial gradients.
We will focus attention on Quintessence models for which the slow-rolling
approximation is valid. In this case, the background satisfies
\bea
\nonumber \dot{Q}&\simeq& -{\frac{V^{\prime }}{3H}}\,,  \label{eom}\\
\nonumber H^2 &\simeq& \frac{8\pi}{3m^2_{pl}}V~,\\
\frac{\dot{H}}{H^2}&\simeq&-\frac{m^2_{pl}}{16\pi}\left(\frac{V^{'}}{V}\right)^2~,
\eea
where $V(Q)$ is the potential for Quintessence field, and a $'$ denotes the
derivative with respect to $Q$. In this case, the expressions for
$\tau_{\mu \nu}$ become

\begin{eqnarray}
\rho _{br} &\equiv& \tau^0_0 = \frac{m_{pl}^2}{8\pi}(12H<\Phi\dot{\Phi}>-3<\dot{\Phi}^2>) \nonumber \\
&+& \frac{1}{2}<\dot{\delta Q}^2>+\frac{1}{2}V''<\delta Q^2>+2V'<\Phi\delta
Q>~,\label{rhobr}\\
\nonumber p_{br} \equiv &-& \frac{1}{3}\tau_i^i=
\frac{m_{pl}^2}{8\pi}(16\dot{H_Q}<\Phi^2>+24H<\Phi\dot{\Phi}>
\nonumber \\
&+& <\dot{\Phi}^2>
+4<\Phi\ddot{\Phi}>+\frac{4V^{'2}}{3V}<\Phi^2>) \nonumber \\
&+& 8V<\Phi^2>+\frac{1}{2}<\dot{\delta Q}^2>-\frac{1}{2}V''<\delta Q^2>
\nonumber \\
&+& 2V'<\Phi\delta Q>+\frac{4V'}{3H}<\dot{\delta
Q}\Phi>~.\label{pbr}
\end{eqnarray}

Each two-point function can be evaluated as an integral over the
Fourier modes (we assume a flat universe for simplicty)
of the linear fluctuation variables. We use the following
convention for the Fourier expansion of
a function U({\bf{x}},t) (e.g., $\Phi$, $\delta Q$):
\begin{equation}\label{u}
U({\bf x},t)=\frac{1}{(2\pi)^{3/2}} \int U_k(t)e^{-i{\bf k\cdot x}}d^3k~.
\end{equation}

Making use of the slow roll approximation for the Quintessence field $Q$,
the linear perturbation equations have the following
solution for $\Phi_{k}$ on scales larger than the Hubble radius
\cite{Mukhanov:1992me}:
\begin{equation}
\Phi_{k} \, \simeq \, A_k \frac{m^2_{pl}}{16\pi}
\left(\frac{V^{'}}{V}\right)^2~ \, ,
\label{phiq}
\end{equation}
where $A_k$ is an integration constant. By taking derivatives of (\ref{phiq})
and using the background equation (\ref{eom}) obtained in the slow-rolling
approximation,we can easily find
\begin{eqnarray}
\dot{\Phi}_{k}
&\simeq&-\frac{2V}{3H_Q}\left[\frac{V^{''}}{V}-\left(\frac{V^{'}}{V}\right)^2\right]\Phi_{k}~,\label{dphiq}\\
\ddot{\Phi}_{k} &\simeq&
\frac{m_{pl}^2}{24\pi}\left(7\frac{V^{'4}}{V^3}+2\frac{V^{'''}V'}{V}-13\frac{V^{'2}V''}{V^2}+4\frac{V^{''2}}{V}\right)\Phi_{k}.\nonumber
\end{eqnarray}

The linearized Einstein constraint equations yield
\begin{eqnarray}
\delta Q_k\, &=& \, \frac{m_{pl}^2}{4\pi
\dot{Q}}(\dot{\Phi}_{k}+H_Q\Phi_{k}) \nonumber \\
&=& \, \left[\frac{m_{pl}^2}{2\pi}\left(\frac{V''}{V'}-\frac{V'}{V}\right)
-2\frac{V}{V'}\right]\Phi_k~,\label{dqk}
\end{eqnarray}
from which the time derivative $\dot{\delta Q}_k$ can easily be expressed
in terms of $\Phi_k$, making use once again of the
background equations in the slow-rolling approximation.

Inserting Eqs.(\ref{u})-(\ref{dqk}) into Eqs. (\ref{rhobr})
and (\ref{pbr}), we obtain:
\begin{eqnarray}
\nonumber \rho_{br} & = &
\{2V(\frac{VV''}{V^{'2}}-2)
+\frac{m_{pl}^2}{12\pi}(\frac{V^{'2}}{V}+10V''-11\frac{VV^{''2}}{V^{'2}})\\
\nonumber &+& \frac{m_{pl}^4}{48\pi^2}(2\frac{V'V'''}{V}-
3\frac{V^{''2}}{V}-4\frac{V^{'2}V''}{V^2}
+3\frac{V^{'4}}{V^3}\\
&-& 2\frac{V''V'''}{V'}+4\frac{V^{''3}}{V^{'2}})
+\frac{m_{pl}^6}{192\pi^3V}(V''' \label{rhobr2} \\
&+& \frac{V^{''2}}{V'}-5\frac{V''V'}{V}+3\frac{V^{'3}}{V^2})^2\}
<\Phi^2> ,  \nonumber \\
\nonumber p_{br} & = &
\{-2V(\frac{VV''}{V^{'2}}-2)
+\frac{m_{pl}^2}{12\pi}(7\frac{V^{'2}}{V}-22V''+13\frac{VV^{''2}}{V^{'2}})\\
\nonumber &+& \frac{m_{pl}^4}{48\pi^2}(25\frac{V^{''2}}{V}
-17\frac{V^{'2}V''}{V^2}+2\frac{V^{'4}}{V^3}-2\frac{V''V'''}{V'}\\
&-& 8\frac{V^{''3}}{V^{'2}})
+\frac{m_{pl}^6}{192\pi^3V}(V''' \label{pbr2} \\
\nonumber &+& \frac{V^{''2}}{V'}-5\frac{V''V'}{V}+3\frac{V^{'3}}{V^2})^2\}
<\Phi^2> .
\end{eqnarray}

The energy density and pressure of the background for
Quintessence during the slow-roll period is
\be
\rho_{bg}\simeq -p_{bg} \simeq V ,
\ee
so we have:
\begin{eqnarray}
\frac{\rho_{br}+3p_{br}}{\rho_{bg}+3p_{bg}} &\simeq& <\Phi^2>
\{2(\frac{VV''}{V^{'2}}-2) \label{ratio} \\
\nonumber &-&\frac{m_{pl}^2}{12\pi}(11\frac{V^{'2}}{V^2}-28\frac{V''}{V}+14
\frac{V^{''2}}{V^{'2}})\\
\nonumber &-&\frac{m_{pl}^4}{96\pi^2}(2\frac{V'V'''}{V^2}
+72\frac{V^{''2}}{V^2}-55\frac{V^{'2}V''}{V^3}\\
\nonumber &+& 9\frac{V^{'4}}{V^4}-8\frac{V''V'''}{V'V}
-20\frac{V^{''3}}{V^{'2}V})\\
\nonumber &-&\frac{m_{pl}^6}{96\pi^3V^2}(V'''+\frac{V^{''2}}{V'}
-5\frac{V''V'}{V}+3\frac{V^{'3}}{V^2})^2\} .
\end{eqnarray}
When this ratio becomes of order unity, the back-reaction effects
begin to dominate. Hence, in the following we will focus on
calculating this ratio.

The crucial quantity to evaluate is the expectation value
$<\Phi^{2}(t)>$. It obtains contributions from all Fourier modes
of $\Phi$:
\begin{equation}\label{phit2}
<\Phi^{2}(t)>=\int_{k_{i}}^{k_{t}}\frac{k^3}{2\pi^2}|\Phi_{k}(t)|^2
\frac{dk}{k}~,
\end{equation}
where $k_i=a_iH_i$  and $k_t=a(t)H(t)$  are infrared  and ultraviolet
cutoffs respectively. The infrared cutoff is given by length scale
above which there are no fluctuations. This is in turn determined
by the cosmological model. If our Universe results from a period
of inflation, then $k_i$ is given by the comoving wave number corresponding
to the Hubble radius at the beginning of inflation. The ultraviolet
cutoff is a more tricky issue. We will return to a discussion of this
point in \cite{LLZ2}. For the moment, let us remark that the phase
space of ultraviolet modes increases more slowly in an accelerating Universe
than the phase space of infrared modes. 

There are two reasonable
choices for the ultraviolet cutoff: either at constant physical or at
constant comoving wavenumber $k$. The first prescription is more physical.
In an exponentially expanding background, the resulting phase space of
ultraviolet modes is constant, whereas the phase space of infrared
modes grows rapidly as the wavelengths of modes are stretched exponentially
to become larger than the Hubble radius. For quintessence models with 
$p > - \rho$, the ultraviolet phase space does grow slowly, but as long
as the equation of state of quintessence does not differ too much from
$p = - \rho$, the growth of the infrared phase space will exceed the
growth of the ultraviolet phase space. The problem with this first way
of setting the ultraviolet cutoff is that it requires the continuous
creation of modes at the ultraviolet cutoff frequency, and this is hard
to reconcile with unitarity. This problem is avoided if the ultraviolet
cutoff is set at constant comoving wavenumber. In this case, the
phase space of ultraviolet modes always decreases in an accelerating
background cosmology because the Hubble radius is decreasing in comoving
coordinates.

We conclude that, 
once the ultraviolet cutoff is set at some time (e.g. the present
time $t_0$), the ultraviolet terms will be well-controlled in all future.
In the following, in evaluating the strength of back-reaction at time $t$,
we will restrict our attention to the effect of infrared modes,
and hence we will only consider the
contribution to (\ref{phit2}) of modes with wavenumber smaller than
the value $k_{H(t)}$ corresponding to the Hubble radius at time $t$.

\section{Spectrum of Cosmological Fluctuations}

In this section we will estimate the strength of back-reaction for
two classes of Quintessence models, assuming that the spectrum of
cosmological fluctuations is normalized by the recent observations
of CMB anisotropies, and that the spectrum is either scale invariant
or given by a slight tilt. We will find that in the former case
back-reaction is always small, whereas it can become large and end
the period of Quintessence domination in the latter case.

To apply the formulas of the previous section we need to relate
the value of $\Phi_k$ during the period of Quintessence domination,
from now on denoted by $\Phi_{Qk}$, to the
corresponding value $\Phi_{mk}$ at a redshift
just before the Quintessence begins to dominate, i.e. in the
matter-dominated phase. For simplicity, we will take this time
to be the present time $t_0$.

It is convenient to consider separately modes which are outside
the Hubble radius today, and modes which are outside the Hubble
radius at the time $t > t_0$, but inside at the present time. For the
former modes we can use
the conservation
\cite{Bardeen:1983qw,Brandenberger:1984tg,Lyth:1985gv}
of the Bardeen parameter
\begin{equation}
\zeta \equiv \frac{2}{3}\frac{\Phi
+H^{-1}\dot{\Phi}}{1+w}+\Phi,
\end{equation}
to relate the values of $\Phi_{Qk}(t)$ and $\Phi_{mk}(t_0)$. Here,
$w = p / \rho$ is a measure of the equation of state of the
background. Thus,
\begin{eqnarray}\label{z_1}
\frac{2}{3}\frac{\Phi _{mk}+H_m^{-1}\dot{\Phi}_{mk}}{1+w_{m}}+\Phi _{mk}
=\frac{2}{3}\frac{\Phi _{Qk}+H_Q^{-1}\dot{\Phi}_{Qk}}{1+w_Q}+\Phi _{Qk}~,
\end{eqnarray}
where the subscripts $m$ and $Q$ denote whether the quantity is evaluated
during the epoch of matter domination or during Quintessence domination.

During the matter dominated era, $w_{m}=0$ and $\dot{\Phi}_{mk}=0$,
so the left hand side of (\ref{z_1}) is $5\Phi_{mk}/3$.
During the slow roll period of Quintessence domination, we have
(from Eqs.(\ref{eom}) and (\ref{dphiq}))
\begin{eqnarray}\label{hdp}
H_Q^{-1}\dot{\Phi}_{k}= -
A_k^{sup} \frac{m^4_{pl}}{(8\pi)^2}\left(\frac{V^{'}}{V}\right)^2
\left[\frac{V^{''}}{V}-\left(\frac{V^{'}}{V}\right)^2\right]~,
\end{eqnarray}
where the subscript``$^{sup}$'' indicates that the mode is  outside the
present Hubble radius. In addition,
\begin{eqnarray}\label{wq}
1+w_Q=\frac{\dot{Q}^2}{V}=
\frac{m^2_{pl}}{24\pi}\left(\frac{V^{'}}{V}\right)^2
\end{eqnarray}.
Combining Eqs.(\ref{phiq}),
 (\ref{z_1}), (\ref{hdp}) and (\ref{wq}), we have
\begin{eqnarray}
\frac{5}{3}\Phi_{mk}= A_k^{sup} (1+5\epsilon-2\eta)~,
\end{eqnarray}
making use of the two slow-roll parameters,
\begin{eqnarray}
\epsilon &\equiv&
\frac{m^2_{pl}}{16\pi}\left(\frac{V^{'}}{V}\right)^2~\ll 1,\\
\eta &\equiv & \frac{m^2_{pl}}{8\pi} \frac{V^{''}}{V}~\ll 1.
\end{eqnarray}
Thus,  we have
\be\label{aksup}
A_k^{sup}\simeq 5\Phi_{mk}/3\sim \Phi_{mk}~.
\ee
We see that the amplitude of $\Phi_k$ during the period of
Quintessence domination is suppressed by the slow-roll parameter $\epsilon$
compared to the corresponding value in the matter-dominated era.

For the modes which are at the present time $t_0$ inside the Hubble
radius, but exit the Hubble radius before time $t$, we use the
fact that while on scales smaller than the Hubble radius the dominant
mode of $\Phi$ scales as
\begin{equation}
\Phi_{Qk} \propto {\dot{Q_k}} \propto {{V'} \over {3 H_Q}} \, ,
\end{equation}
(see Eq. (6.54) of \cite{Mukhanov:1992me}) while on scales larger
than the Hubble radius
\begin{equation}
\Phi_{Qk} \propto ({{V'} \over V})^2,
\end{equation}
(see (\ref{phiq})). This allows us to express $\Phi_{Qk}$ in terms of the
values of the potential and its derivatives at the times $t_0$, $t_H(k)$
and $t$, where $t_H(k)$ is the time when the scale $k$ crosses the
Hubble radius, and is given by $a(t_H(k)) H = k$.
\bea
\nonumber \Phi_{Qk} &=& \frac{1}{6\sqrt{2\pi}m_{pl}}
\left(\frac{V^{3/2}}{V'}\right)_{t_H(k)}\frac{V^{'2}}{V^2}
\Phi_{mk}\left(\frac{1}{6\sqrt{2\pi}m_{pl}}\frac{V_0^{'}}{V_0^{1/2}}\right)^{-1}\\
&=& \Phi_{mk}\frac{V_0^{1/2}}{V_0^{'}}\left(\frac{V^{3/2}}{V'}\right)_{t_H(k)}
\frac{V^{'2}}{V^2} \nonumber \\
&\equiv& A_k^{sub}\frac{m^2_{pl}}{16\pi}\frac{V^{'2}}{V^2}~,
\eea
where
\be\label{aksub}
A_k^{sub}\equiv \frac{16\pi}{m_{pl}^2}
\frac{V_0^{1/2}}{V_0^{'}}\left(\frac{V^{3/2}}{V'}\right)_{t_H(k)}\Phi_{mk}~
\ee
is introduced to accord with the notation of (\ref{phiq}). Note that since
the Quintessence field is a scalar field, we were able to use equations
for cosmological perturbations derived for a scalar field-dominated
equation of state.

In the following, we will evaluate the two point function (\ref{phit2}) for
times $t \gg t_0$, making use of the amplitudes of $\Phi_{Qk}$ derived
above. We will consider two potentials commonly used for Quintessence.
First, we will assume that the spectrum of fluctuations at the present
time is scale-invariant. Later, we will relax this assumption and consider
blue spectra.

\subsection{Scale Invariant Spectrum}

If the present spectrum of fluctuations is scale-invariant, then
\begin{equation}\label{sip}
P_m(k) \equiv \frac{k^3}{2\pi^2}|\Phi_{mk}|^2= C \, ,
\end{equation}
where the constant $C$ can be fixed by the amplitude of CMB anisotropies
on large angular scales measured by COBE \cite{Smoot:2000si}
\begin{equation}\label{cobe}
\delta_H\simeq 10^{-5}\simeq P_m(k)^{1/2}_{k=aH_m}=C^{1/2} \, .
\end{equation}

Substituting Eqs.(\ref{phiq}), (\ref{aksup}), (\ref{aksub}),
(\ref{sip}) and (\ref{cobe}) into (\ref{phit2}), we finally arrive at:
\begin{eqnarray} \label{phiq2}
\nonumber <\Phi_Q ^{2}(t)> &\simeq &
\int_{k_i}^{k_0}C\left(\frac{m_{pl}^2}{16\pi}\right)^2\frac{V^{'4}}{V^4}\frac{dk}{k} \\
&+& \int_{k_0}^{k_t}C\frac{V_0}{V_0^{'2}}
\left(\frac{V^3}{V^{'2}}\right)_{t_{k=aH_Q}}\frac{V^{'4}}{V^4}\frac{dk}{k}\\
\nonumber &=& C \epsilon^2 \ln \frac{a_0H_0}{a_iH_i} \\
\nonumber &+&
C \frac{V_0}{V_0^{'2}}\frac{V^{'4}}{V^4}
\int_{k_0}^{k_t}\left(\frac{V^3}{V^{'2}}\right)_{t_H(k)}d\ln k \, .
\end{eqnarray}
Replacing the integral over $k$ by the integral over $a(t_H(k))$ we obtain
\begin{equation}
<\Phi_Q ^{2}(t)> \simeq  C \epsilon^2 \ln \frac{a_0H_0}{a_iH_i} +
C \frac{V_0}{V_0^{'2}}\frac{V^{'4}}{V^4}
\int_{a_0}^{a_t}\left(\frac{V^3}{V^{'2}}\right)d\ln a \, , \label{phi2fin}
\end{equation}
where we have neglected $d\ln H_Q = -\epsilon d \ln a$.

The slow-rolling approximation gives:
\begin{eqnarray}
\nonumber a(t) &=&
a_{0}\exp \left[-\frac{8\pi}{m_{pl}^2}
\int_{Q_{0}}^{Q}\frac{V}{V^{^{\prime}}}dQ\right] \, ,\\
d \ln a &=& -\frac{8\pi }{m_{pl}^2}\frac{V}{V'}d Q \, .
\end{eqnarray}
Using these relations, the expression (\ref{phi2fin}) can be simplified,
yielding
\begin{eqnarray}  \label{Phiq2t}
\nonumber
<\Phi^2_Q(t)>
&\simeq & C \epsilon^2 \ln \frac{a_0H_0}{a_iH_i} \\
\nonumber &-&
C \frac{8\pi}{m_{pl}^2} \frac{V_0}{V_0^{'2}}\frac{V^{'4}}{V^4}
\int_{Q_0}^{Q}\left(\frac{V^4}{V^{'3}}\right)d Q~.
\end{eqnarray}

The first of the two Quintessence models for which we will evaluate
the effect of back-reaction is
\begin{equation} \label{pot1}
V = B Q^{-\alpha}
\end{equation}
with $B$ being a constant and $\alpha>1$. First, note that for
this potential, the first term in (\ref{rhobr2}) and (\ref{pbr2})
dominates and determines the equation of state of the effective
energy-momentum tensor of back-reaction. In particular, it
follows that the equation of state is that of a {\it negative} cosmological
constant, as in the case of the back-reaction in inflationary
cosmology \cite{Mukhanov:1997ak,Abramo:1997hu}.
Thus, we see that the effects of back-reaction
oppose the Quintessence-driven acceleration. Inserting (\ref{pot1}) into (\ref{phiq2})
yields
\begin{eqnarray}\label{alpha4}
<\Phi^2_Q(t)> &=&
C\frac{1}{\pi^2}\frac{m_{pl}^4}{Q^4}\ln \frac{a_0H_0}{a_iH_i} \\
\nonumber &+&
C\frac{2\pi}{m_{pl}^2}\frac{Q_0^6}{Q^4}\ln \frac{Q}{Q_0} ,
~~~~ {\rm for}~~ \alpha=4~,
\end{eqnarray}
\begin{eqnarray}\label{alphan4}
<\Phi^2_Q(t)> &=&
C \frac{\alpha^4}{256\pi^2}\frac{m_{pl}^4}{Q^4}\ln \frac{a_0H_0}{a_iH_i} \\
\nonumber &+&
C \frac{8\pi}{\alpha (4-\alpha)}\frac{Q_0^2}{m_{pl}^2}
\left(\frac{Q_0^{\alpha}}{Q^{\alpha}}-\frac{Q_0^4}{Q^4}\right) ,
~~~~ {\rm for}~~ \alpha \neq 4~,
\end{eqnarray}
where $Q_0 = Q(t_0)$.
Substituting (\ref{pot1}) into (\ref{ratio}), we have
\begin{eqnarray}  \label{con}
\frac{\rho_{br}+3p_{br}}{\rho_{bg}+3p_{bg}} & &
\simeq \\
\nonumber & &
\left[2\left(\frac{1}{\alpha}-1\right)+
O\left(\left(\frac{m_{pl}}{Q}\right)^2\right)\right]<\Phi^2_Q(t)>
\end{eqnarray}
In this model, $Q_0\sim m_{pl}$, and $Q$ increases with time.
It can be seen from the combination of
Eqs. (\ref{alpha4}), (\ref{alphan4}) and (\ref{con}) that
the backreaction is very small now and will be smaller in the future.
Notice that the contribution of the first term in
(\ref{alpha4}) and (\ref{alphan4})
is generally small in this case. For example,
if we assume the inflation started at GUT scale
and e-folding number is 70, $\ln  \frac{a_0H_0}{k_i}$ is about
$O(10^1)$. When $Q>> m_{pl}$, the first term can be neglected
compared with the second term in Eqs. (\ref{alpha4}) and
(\ref{alphan4}).

The second model is
\begin{equation} \label{pot2}
V \, = \, D e^{-\lambda Q /m_{pl}}
\end{equation}
with $D$ and $\lambda$ constants. In this model,
$\epsilon=\lambda^2/16\pi\ll 1$,
$\eta=\lambda^2/8\pi\ll 1$.
It can easily be verified that also for this potential the first term on
the right hand side of (\ref{rhobr2}) and (\ref{pbr2}) dominates, and
once again leads to an equation of state of the effective energy-momentum
tensor of back-reaction which acts like a negative cosmological constant
and hence opposes the Quintessence-driven acceleration.
After some straightforward calculation we have
\begin{eqnarray}  \label{con2}
\frac{\rho_{br}+3p_{br}}{\rho_{bg}+3p_{bg}}
&\simeq&
-C\left(2+\frac{\lambda^2}{4\pi}\right) \\
\nonumber & & \left[\frac{\lambda^4}{256\pi^2}
\ln \frac{a_0H_0}{a_iH_i} +
\frac{8\pi}{\lambda^2}\left(1-e^{-\lambda(Q-Q_0)/m_{pl}}\right)\right]\\
&\simeq &
-C\frac{16\pi}{\lambda^2}\left(1-e^{-\lambda(Q-Q_0)/m_{pl}}\right)\
\nonumber.
\end{eqnarray}
One can see that for a reasonable value of $\lambda$\cite{Ferreira:1997au},
the backreaction in this model is very small too.

\subsection{Blue Perturbation Spectra}

In the above, we assumed that the spectrum for the gravitational
potential during the matter dominated
era $P_m(k)$ is exactly scale invariant.
However, if the primordial spectrum has a slight blue tilt,
the conclusion will change dramatically.
Thus, we now assume that
\be
P_{m}(k) \, = \, C\left(\frac{k}{k_{COBE}}\right)^n,
\ee
with $0 <n < 0.1$, the upper bound being set by the joint analysis
of \cite{Jaffe:2001tx} of the Maxima-1, Boomerang and COBE cosmic microwave
anisotropy results. The upper bound corresponds to the one sigma statistical
error. Including the estimate of \cite{Jaffe:2001tx} of the systematic
errors would increase the bound to $n < 0.27$.
In the above, $k_{COBE}\simeq 7.5 a_0H_0$.
The constant $C$ is again given by (\ref{cobe}).
Straightforward calculations yield
\begin{eqnarray}
& & <\Phi^2_Q(t)>
\simeq  \frac{C \epsilon^2}{7.5^nn}
\left(1-\left(\frac{a_iH_i}{a_0H_0}\right)^n\right) \\
\nonumber &-& \frac{C}{7.5^n} \frac{8\pi}{m_{pl}^2}
\frac{V_0^{1-n/2}}{V_0^{'2}}
\frac{V^{'4}}{V^4}\int_{Q_0}^{Q}\frac{V^{4+n/2}}{V^{'3}}
\left(e^{-\frac{8\pi n}{m_{pl}^2}\int_{Q_0}^{Q}\frac{V}{V'}dQ}\right)
d Q \, .
\end{eqnarray}
As argued in the above, the first term is generally small
(notice $n$ is small which also implies that $7.5^n\simeq 1$).

For the first model (\ref{pot1}), we have
\begin{eqnarray} \label{mo1}
& & <\Phi^2_Q(t)> \simeq \\
\nonumber & &
C\left(\frac{4\pi (-n)}{\alpha m_{pl}^2}\right)^{-1+\frac{(1+n/2)\alpha}{2}}
{1 \over {-n}} Q_0^{2+(1+n/2)\alpha}
e^{-4\pi n Q_0^2/\alpha m_{pl}^2}\\
\nonumber & &
Q^{-4}
\Gamma\left(2-\frac{(1+n/2)\alpha}{2},-\frac{4\pi n}{\alpha m_{pl}^2}Q_0^2,
-\frac{4\pi n}{\alpha m_{pl}^2}Q^2\right)~,
\end{eqnarray}
where $\Gamma(a, z_0, z_1)$ is the generalized incomplete gamma
function:
\begin{eqnarray} \label{gamma}
\Gamma(a,z_0,z_1) &=& \int_{z_0}^{z_1}x^{a-1}e^{-x}dx\\
\nonumber &=&
\frac{e^{-z_0}z_0^a}{\Gamma(1-a)}\int_0^{\infty}\frac{e^{-x}x^{-a}}
{z_0+x}dx \\
\nonumber &-& \frac{e^{-z_1}z_1^a}{\Gamma(1-a)}\int_0^{\infty}\frac{e^{-x}x^{-a}}
{z_1+x}dx~.
\end{eqnarray}
From Eqs.(\ref{mo1}) and (\ref{gamma}) we can see that
the back-reaction will become important and terminate the acceleration
of the Universe driven by the background of Quintessence
when $Q$ becomes enough large, because we approximately have
\begin{equation}
<\Phi^2_Q(t)> \, \propto \, \frac{e^{4\pi n Q^2/\alpha m_{pl}^2}}
{Q^{2+(1+n/2)\alpha}} \, .
\end{equation}

For the second model(\ref{pot2}) we have
\begin{equation}
<\Phi^2_Q(t)> \, \simeq \, C\frac{8\pi}{\lambda
m_{pl}}(Q-Q_0) \, ,\label{mo21}
\end{equation}
when $1+n(\frac{1}{2}-\frac{1}{\eta})=0$~, and
\begin{eqnarray}
& & <\Phi^2_Q(t)> \, \simeq \\
\nonumber & & \frac{C/\eta}{1+n(\frac{1}{2}-\frac{1}{\eta})}
\left(1-e^{-\frac{\lambda}{m_{pl}}(1+n(\frac{1}{2}-\frac{1}{\eta}))
(Q-Q_0)}\right)~.\label{mo22}
\end{eqnarray}
when $1+n(\frac{1}{2}-\frac{1}{\eta})\ne 0$~.
In the first case, the back-reaction will terminate the
acceleration of the Universe when
$Q\sim 10^{10}\sqrt{\eta/8\pi}m_{pl}$. In the second case, for
$n>2\eta/(2-\eta)$
the acceleration will stop when
\be
Q\sim \sqrt{\frac{\eta}{8\pi}}\frac{\ln [10^{10}(n-(1+n/2)\eta)+1]}
{n-(1+n/2)\eta}{m_{pl}}+{Q_0}.
\ee

Recall that the slow rolling parameter $\eta$ refers to the dynamics of
quintessence whereas the index $n$ refers to the spectrum of fluctuations
produced during inflation. Therefore, $n$ and $\eta$ are independent, and
our constraint $n > 2 \eta / (2 - \eta)$ for back-reaction to be important
does not represent a severe fine-tuning.

Note that for $n<2\eta/(2-\eta)$, similar to Eq. (\ref{con2}), 
the backreaction is generally still
small unless the difference between the parameters $n$ and $\eta$
is fine tuned a lot, for example, if
\begin{eqnarray}
\frac{\eta-10^{-10}}{1-\eta/2} <n< \frac{\eta}{1-\eta/2},
\end{eqnarray}
the contribution coming from the back-reaction will terminate the
accelerating expansion of the Universe in the distant future.

\section{Discussion}

We have demonstrated that the gravitational back-reaction of cosmological
perturbations can lead to a termination of the period of Quintessence
domination, in the same way it can lead to a termination of a period
of inflation \cite{Mukhanov:1997ak,Abramo:1997hu}.
In this note we have considered the
effect of infrared modes, i.e. modes whose wavelength at time $t$ (when
back-reaction is evaluated) is larger than the Hubble radius.

Specifically, we assume a COBE normalized spectrum of cosmological
fluctuations at the present time $t_0$. We consider both a scale-invariant
spectrum, and spectra with slight blue tilts. We study two classes of
Quintessence potentials, given by (\ref{pot1}) and (\ref{pot2}). The
most important contribution to back-reaction comes from modes with
$k_0 < k < k_H(t)$, where $k_0$ is the wave number corresponding to the
present Hubble radius, and $k_H(t)$ that corresponding to the Hubble
radius at time $t$. If the spectrum has a blue tilt, it is the
upper limit which dominates.
We find that for spectra with a sufficiently large blue tilt, back-reaction
will terminate the period of Quintessence domination. The effective
energy-momentum tensor which describes the back-reaction has the form
of a negative cosmological constant and hence opposes the Quintessence-driven
acceleration.

Since the dominant contribution in our analysis comes from the largest
values of $k$ we consider, it is to be expected that the contribution
to back-reaction from ultraviolet modes (modes whose wavelength at time $t$
is smaller than the Hubble radius), may dominate over the effects
calculated here. We will return to this issue in \cite{LLZ2}. The
point of our analysis was to show that back-reaction cannot be neglected
when considering the future of a Universe dominated at the present
time by Quintessence.

The wavelength of modes which dominate the contribution to our effect was
smaller than the Hubble radius at $t_0$. Hence, these modes are subject
to the usual nonlinear growth of cosmological fluctuations in the matter
dominated phase. This is another effect which has not been taken into
account here (but will be in \cite{LLZ2}). However, it is reasonable
to expect that nonlinear effects will enhance the amplitude of the
fluctuations and hence increase the back-reaction effects.

Obviously, if the spectrum of fluctuations is so close to scale-invariance
so that back-reaction only become important in the very distant future,
then the modes which will contribute in this case had a microphysical
wavelength at the present time, and we are faced with the true ultraviolet
problem of field theory.

It has recently been claimed \cite{Fischler:2001yj,Hellerman:2001yi}
that the presence of future horizons in
Quintessence models poses a problem in trying to reconcile string theory
and Quintessence. In models where our back-reaction effect becomes
important, this problem may well disappear (without the need to have
to introduce extra physics such as additional scalar fields).

\vspace{0.5cm}
\centerline{\bf Acknowledgments}
\vspace{0.2cm}

One of us (R.B.) wishes to thank the Institute of High Energy Physics in
Beijing for their wonderful hospitality during the time when this work
was started. The research was supported in
part by the U.S. Department of Energy under Contract DE-FG02-91ER40688, TASK A
and by the NSF of China and the Ministry of Science and Technology
of China under Grant No.NKBRSF G19990754.



\end{document}